\documentclass[preprint,12pt]{elsarticle}

\usepackage[T1]{fontenc}
\usepackage{amsmath,amssymb}
\usepackage{physics}
\usepackage{graphicx}

\usepackage{xurl}
\providecommand{\href}[2]{#2}
\journal{Results in Optics}
\biboptions{numbers,sort&compress}

\begin{document}

\begin{frontmatter}

\title{Partial-coherence weak measurement of polarization qubits with a
temporal pointer: theory and experiment}

\author[moncton]{S. P. Gervais\fnref{fn1}}
\author[moncton]{L. Giner\corref{cor1}}
\ead{lambert.giner@umoncton.ca}

\affiliation[moncton]{organization={D\'epartement de physique et d'astronomie,
                                     Universit\'e de Moncton},
                      city={Moncton},
                      state={New Brunswick},
                      country={Canada}}

\fntext[fn1]{Present address: Department of Astronomy \& Physics, Saint Mary's University,
Halifax, Nova Scotia, Canada.}
\cortext[cor1]{Corresponding author.}

\begin{abstract}
Weak measurement has proven useful for metrology, for probing quantum paradoxes,
and for the direct measurement of quantum states. Weak measurements of
polarization qubits with temporal pointers provide a direct route to weak-value
characterization, including in regimes where the pointer is only partially
coherent. We derive a closed-form expression for the post-selected temporal
shift that interpolates between the incoherent and coherent (ideal)
weak-measurement limits through a single parameter $\nu$, the pointer
visibility. The expression is experimentally tested for a range of polarization
states spanning three trajectories corresponding to rotations about the
principal axes of the Poincar\'e sphere. Two of the trajectories provide
parameter-free benchmarks, with the real part of the weak value following
$\cos^2\theta$ and $1/2$. In contrast, the weak value along the third trajectory
is sensitive to incoherence and is well described by the single-parameter
partial-coherence model, from which a reduced pointer visibility is obtained.
Time shifts are extracted from the smoothed derivative of the pulse leading edge
with internal calibration, reaching few-picosecond precision on
$10~\mathrm{ns}$ optical pulses using commercial components. The same apparatus
also gives qualitative access to the imaginary quadrature through the frequency
centroid, showing that temporal weak measurements remain accessible and
quantitative in partially coherent pulsed systems.
\end{abstract}


\end{frontmatter}

\section{Introduction}
\label{sec:intro}

Weak measurement was introduced by Aharonov, Albert, and Vaidman as a protocol
in which a preselected quantum system is weakly coupled to a pointer and then
post-selected, revealing an observable quantity known as the weak value \cite{Aharonov}. The
weak value is generally complex and can lie outside the eigenvalue spectrum of
the measured observable, a feature that gives weak measurements their
metrological interest through the amplification of small physical shifts
\cite{Lundeen_Bamber,ApplicationWeak,Btwn_pre_post,JordanBook}, with recent
schemes showing that real weak values in particular can enhance interferometric
sensitivity \cite{intlundeen}. The real and imaginary parts of the weak value appear in conjugate pointer
quadratures, such as position and momentum in spatial implementations or time
and frequency in temporal implementations. Weak-measurement ideas have been applied across
single-photon polarization measurements, beam-deflection metrology, field
measurements, and related optical platforms \cite{troupe2017quantumcryptographyweakmeasurements,WeakorStd,OpticalNetworks,Jeff_outperform,Magaña-Loaiza_2017,SpinHallEffect,Aharonov_2002_Hardy,HardyParadox,Lundeen_Resch,imaginary_part,QuantumLimitedMetrology, RydbergDispersion,Huang:22}. The practical implementation,
however, depends strongly on the choice and quality of the pointer.

Most optical implementations use spatial pointers, where the measured quantity
is inferred from a beam displacement \cite{Guilleaum,Lundeen_Direct_Measurement,weak_photon_pol}. Spectral and temporal pointers use the
same weak-coupling logic but replace spatial imaging with frequency or
time-resolved detection \cite{OpticalNetworks,Howell2022,Time_delay_china,quantuminfoweak}. In the temporal and spectral domain, this
approach was first used to measure group and signal velocities in optical
fiber~\cite{Brunner_2004}, and later combined with spectral interference to
resolve temporal delays well below the pulse width~\cite{Salazar}. Beyond the
real part, the real and imaginary quadratures of the weak value have been read
out simultaneously in polarization implementations~\cite{Hairiri}, which
motivates the temporal-pointer measurements reported here, where both
quadratures are extracted from each recorded pulse, the real one
quantitatively and the imaginary one qualitatively. Temporal pointers are attractive for polarization
qubits because they can be implemented with standard pulsed lasers and fast
electronic detection, without specialized spatial or spectral imaging. The
trade-off is that temporal precision is bounded by pulse coherence, and
inexpensive pulsed sources can be only partially
coherent on the delay scales relevant to weak measurement conducted with typical
mechanisms such as birefringent walk-off. A single partial-coherence parameter has been
used for a spatial pointer with pseudo-thermal light~\cite{Cho2010}; here it
arises intrinsically from a commercial multimode pulsed source in the temporal
domain. This partial pointer coherence is not a failure mode of the experiment. It
shifts the measured signal in a controlled way, and is therefore included
explicitly in the model.

In this work, we derive a closed-form expression for the post-selected time
shift that interpolates between the projective $\nu=0$ and ideal weak
$\nu=1$ limits, where $\nu$ is the pointer visibility at the
measurement delay. We then use this expression to choose three input-state
trajectories corresponding to the principal axes of the Poincar\'e sphere. Two of these
trajectories have $\nu$-independent predictions and provide parameter-free
checks of the apparatus and extraction method, while the third depends
explicitly on $\nu$ and constrains the visibility of the partially coherent
pointer. This internal
consistency across complementary trajectories is the main
experimental test of the partial-coherence framework.

The experiment uses a commercial multimode pulsed diode laser and a fast
oscilloscope, with time shifts extracted from the smoothed derivative of the
pulse leading edge and calibrated internally in each acquisition. This procedure reaches few-picosecond timing precision on $10~\mathrm{ns}$
pulses, sufficient to resolve the visibility-dependent time shifts in the
partial-coherence regime considered here. Section
\ref{sec:theory} presents the theory and the three Poincar\'e-sphere
trajectories, Sec. \ref{sec:setup} describes the experimental setup, Sec.
\ref{sec:method} details the time-shift extraction method, Sec.
\ref{sec:results} reports the real- and imaginary-quadrature measurements, and
Sec. \ref{sec:discussion} discusses the limitations and conclusions.

\section{Theory}
\label{sec:theory}

We first set up the weak-measurement protocol with post-selection on $\ket{D}$,
then derive the post-selected time shift for a partially coherent pointer, and
finally specialize it to the three Poincar\'e-sphere trajectories used in the
experiment.

\subsection{Weak measurement with post-selection on $\ket{D}$}
Let the initial polarization state be
$\ket{\psi}=a\ket{H}+b\ket{V}$, where $a,b\in\mathbb{C}$ and
$|a|^2+|b|^2=1$. The temporal pointer has amplitude $\xi(t)$, is normalized as
$\int |\xi(t)|^2\,dt=1$, and is centered at $t=0$. The weak interaction is
described by the von Neumann operator
\begin{equation}
\hat U=\exp\!\left(-i\tau\,\hat\Pi_H\otimes\hat p_t\right),
\qquad
\hat\Pi_H=\ket{H}\bra{H},
\label{eq:vonneumann}
\end{equation}
where $\tau$ is the delay applied to the $\ket{H}$ component and $\hat p_t = -i\,\partial/\partial t$ is conjugate to time
($\hbar = 1$). This interaction shifts
the horizontal component of the pointer by $\tau$ and leaves the vertical
component unchanged. Post-selection on
$\ket{D}=(\ket{H}+\ket{V})/\sqrt{2}$ gives the exact conditional pointer
amplitude
\begin{equation}
F(t)=\frac{a}{\sqrt{2}}\,\xi(t-\tau)+\frac{b}{\sqrt{2}}\,\xi(t).
\label{eq:Ft}
\end{equation}
The weak value of the projector is~\cite{Lundeen_Direct_Measurement}
\begin{equation}
W=\frac{\bra{D}\hat\Pi_H\ket{\psi}}{\braket{D}{\psi}}
=\frac{a}{a+b}.
\label{eq:weakvalue}
\end{equation}
The real and imaginary parts of $W$ are accessed through the temporal and
spectral quadratures of the pointer ~\cite{Lundeen_Resch,Jozsa2007}. In the weak regime, $\tau\ll\sigma$, where
$\sigma$ is the effective temporal width of the pointer, the centroid of
$|F(t)|^2$ reduces to $\tau\,\mathrm{Re}(W)$; in the usual weak Gaussian-pointer
limit the conjugate frequency shift scales as
\begin{equation}
\langle\hat\omega\rangle \simeq 2\,\mathrm{Im}(W)\,\tau\,\mathrm{Var}(\omega),
\label{eq:omega_shift}
\end{equation}
where $\mathrm{Var}(\omega)$ is the spectral variance of the pointer~\cite{Jozsa2007}. Beyond the linear weak-coupling limit, the temporal displacement depends on the
overlap between $\xi(t)$ and $\xi(t-\tau)$, which is the role of the visibility
parameter introduced below.

\subsection{Time shift with partial pointer coherence}
The measured time shift is the centroid of the pointer probability distribution,
\begin{equation}
\langle\hat t\rangle
=\frac{\int t\,|F(t)|^2\,dt}{\int |F(t)|^2\,dt}.
\label{eq:centroid}
\end{equation}
For compactness, define $A=a/\sqrt{2}$ and $B=b/\sqrt{2}$, so that
$F(t)=A\xi(t-\tau)+B\xi(t)$. Substituting Eq. \eqref{eq:Ft} into
Eq. \eqref{eq:centroid} gives terms proportional to
$|A|^2|\xi(t-\tau)|^2$, $|B|^2|\xi(t)|^2$, and the interference term
$2\,\mathrm{Re}\!\left[A\bar B\,\xi(t-\tau)\xi^*(t)\right]$. Using the centering
conditions $\int t|\xi(t)|^2\,dt=0$ and
$\int t|\xi(t-\tau)|^2\,dt=\tau$, the real-part overlap at the delay is written
as
\begin{equation}
\nu=
\frac{\mathrm{Re}\!\int \xi^*(t)\,\xi(t-\tau)\,dt}
     {\int |\xi(t)|^2\,dt}.
\label{eq:visibility}
\end{equation}
The quantity $\nu$ lies between 0 and 1 and captures the pointer
coherence relevant to the chosen delay. Under the same overlap approximation
used to express the cross-centroid term, the observable time shift becomes
\begin{equation}
\langle\hat t\rangle
=\tau\,
\frac{|a|^2+\nu\,\mathrm{Re}(a\bar b)}
     {1+2\nu\,\mathrm{Re}(a\bar b)}.
\label{eq:tmean}
\end{equation}
The left-hand side of Eq.~\eqref{eq:tmean} defines the post-selected normalized
pointer shift, $\widetilde{W}\equiv\langle\hat t\rangle/\tau$. For $\nu=1$, Eq. \eqref{eq:tmean} gives 
the coherent weak-measurement response and reduces to $\tau\,\mathrm{Re}(W)$ in the weak regime. For
$\nu=0$, it gives the projective result
$\langle\hat t\rangle=\tau |a|^2$, corresponding to the Born-rule probability
for $\hat\Pi_H$. Intermediate values of $\nu$ interpolate between these
two limits. The visibility dependence is controlled by
$\mathrm{Re}(a\bar b)$, which is set by the trajectory followed on the
Poincar\'e sphere. Equation~\eqref{eq:tmean} follows from the exact centroid of $|F(t)|^2$ and
assumes only normalization, centering, and the real overlap $\nu$; it does not
rely on an expansion of $\xi(t)$ in powers of $\tau$. The near-rectangular
pulse used here therefore falls outside the usual weak-coupling expansions,
while remaining within the scope of Eq.~\eqref{eq:tmean}.

\subsection{Three trajectories on the Poincaré sphere}
\begin{figure}[htbp]
\centering
\includegraphics[width=\columnwidth]{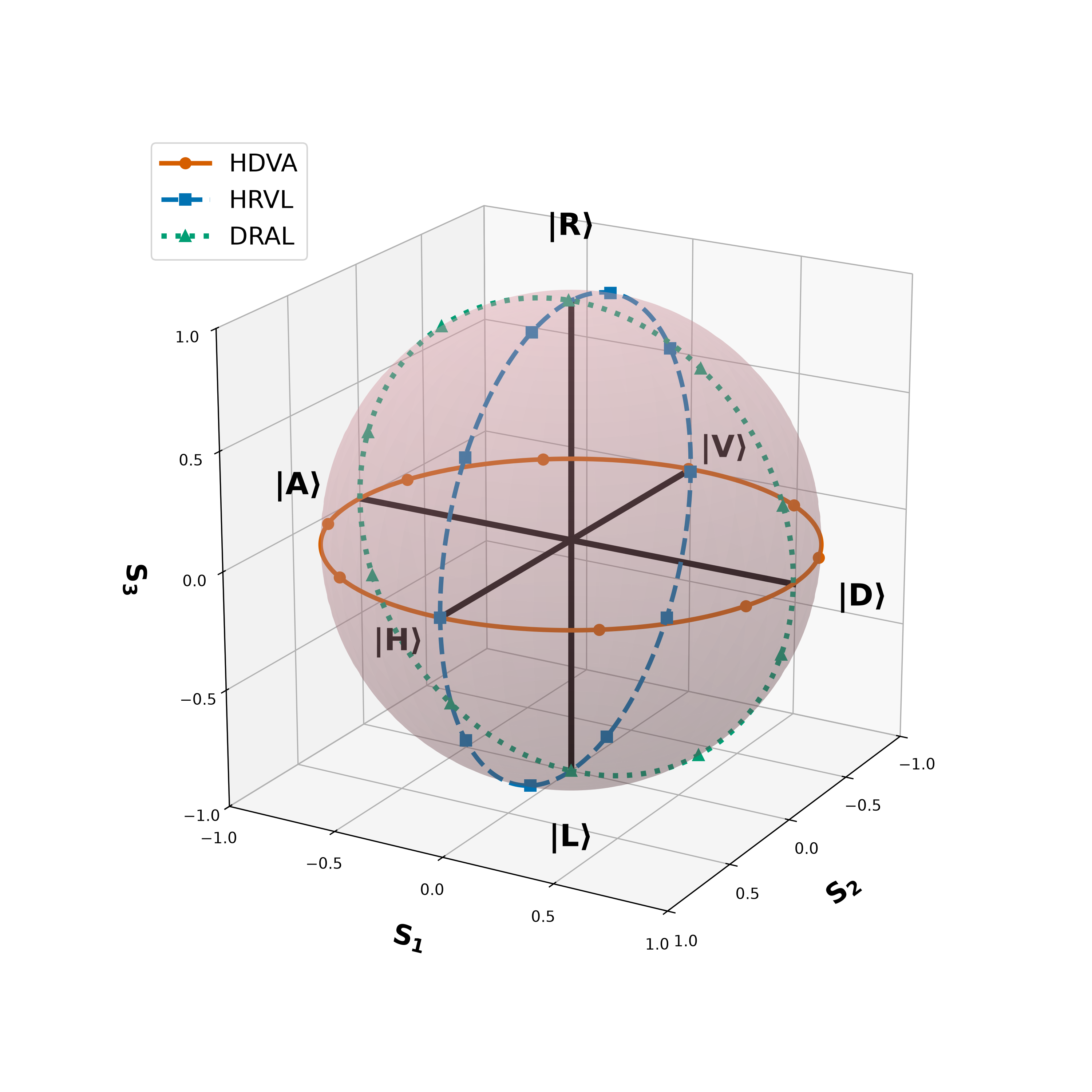}
\caption{Poincar\'e-sphere representation of the three polarization
trajectories. The HDVA (orange, solid, circles), HRVL (blue, dashed, squares),
and DRAL (green, dotted, triangles) paths are rotations about $S_3$, $S_2$, and
$S_1$, and pass respectively through
$\{\ket{H},\ket{D},\ket{V},\ket{A}\}$, $\{\ket{H},\ket{R},\ket{V},\ket{L}\}$,
and $\{\ket{D},\ket{R},\ket{A},\ket{L}\}$.}
\label{fig:poincare_paths}
\end{figure}

The three trajectories below are obtained by standard Jones-matrix
calculations. They are rotations about the principal axes of the Poincar\'e
sphere and were chosen because they span distinct responses to pointer
coherence while retaining compact analytic predictions. In simple terms, the
Stokes parameters are the three coordinates of a polarization state on the
Poincar\'e sphere: $S_1$ distinguishes horizontal from vertical polarization,
$S_2$ distinguishes diagonal from antidiagonal polarization, and $S_3$
distinguishes right- from left-circular polarization. Throughout this section,
$\theta$ denotes the polarization-state angle along the chosen trajectory; in
the experiment it is obtained from the HWP scan as
$\theta=2\theta_{\mathrm{pol}}$, where $\theta_{\mathrm{pol}}$ is the HWP angle
relative to the trajectory origin. The
horizontal-diagonal-vertical-antidiagonal (HDVA) trajectory is a rotation about
$S_3$, generated by an HWP acting on
$\ket{H}$:
\begin{equation}
\ket{\psi_{\mathrm{HDVA}}}
=\cos\theta\ket{H}+\sin\theta\ket{V}.
\label{eq:state_hdva}
\end{equation}
It passes through $\ket{H}$, $\ket{D}$, $\ket{V}$, and $\ket{A}$. The
horizontal-right-vertical-left (HRVL) trajectory is a rotation about $S_2$,
generated by an HWP followed by a quarter-wave plate (QWP) set to
$\theta_{\mathrm{pol}}=0^\circ$:
\begin{equation}
\ket{\psi_{\mathrm{HRVL}}}
=\mathrm{e}^{-i\pi/4}\left(
\cos\theta\ket{H}+i\sin\theta\ket{V}\right),
\label{eq:state_hrvl}
\end{equation}
and passes through $\ket{H}$, $\ket{R}$, $\ket{V}$, and $\ket{L}$. The
diagonal-right-antidiagonal-left (DRAL) trajectory is a rotation about $S_1$,
generated by an HWP followed by a QWP set to $\theta_{\mathrm{pol}}=45^\circ$:
\begin{equation}
\ket{\psi_{\mathrm{DRAL}}}
=\frac{1}{\sqrt{2}}\left(\ket{H}+\mathrm{e}^{2i\theta}\ket{V}\right).
\label{eq:state_dral}
\end{equation}
This trajectory has $|a|=|b|=1/\sqrt{2}$ and passes through $\ket{D}$,
$\ket{R}$, $\ket{A}$, and $\ket{L}$.
The polarization paths defined by Eqs. \eqref{eq:state_hdva}--\eqref{eq:state_dral}
are plotted on the Poincar\'e sphere in Fig. \ref{fig:poincare_paths}.

\begin{figure}[htbp]
\centering
\includegraphics[width=\columnwidth]{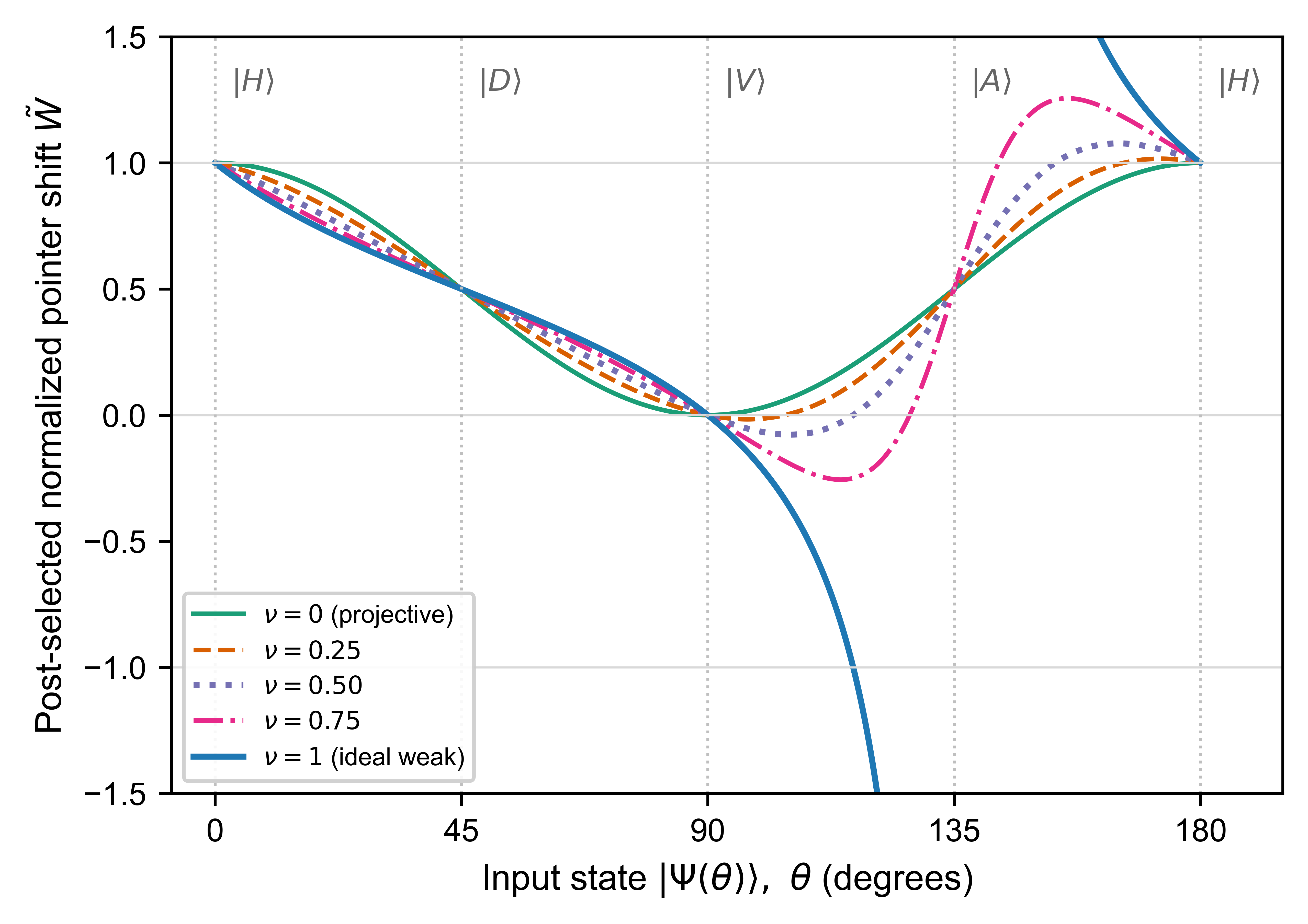}
\caption{Theoretical prediction for the post-selected normalized pointer shift
$\widetilde{W}$ along the HDVA trajectory as
a function of the polarization-state angle $\theta$, shown for several pointer
visibilities $\nu$. The projective limit
$\nu=0$ reduces to $\cos^2\theta$, while the ideal weak-value limit
$\nu=1$ diverges near $\ket{A}$ at $\theta=135^\circ$. Intermediate
visibilities interpolate between these limits and remain bounded over most of
the trajectory.}
\label{fig:theory_hdva}
\end{figure}

For HDVA, HRVL, and DRAL, respectively,
$\mathrm{Re}(a\bar b)=\tfrac12\sin(2\theta),\ 0,\ \tfrac12\cos(2\theta)$,
with $|a|^2=\cos^2\theta,\ \cos^2\theta,\ \tfrac12$.
Substituting into Eq.\eqref{eq:tmean} gives

\begin{equation}
\widetilde{W}_{\mathrm{HDVA}}
=\frac{\langle\hat t\rangle_{\mathrm{HDVA}}}{\tau}
=\frac{\cos^2\theta+\nu\sin\theta\cos\theta}{1+\nu\sin(2\theta)},
\label{eq:traj_hdva}
\end{equation}
\begin{equation}
\widetilde{W}_{\mathrm{HRVL}}
=\frac{\langle\hat t\rangle_{\mathrm{HRVL}}}{\tau}
=\cos^2\theta,
\label{eq:traj_hrvl}
\end{equation}
and
\begin{equation}
\widetilde{W}_{\mathrm{DRAL}}
=\frac{\langle\hat t\rangle_{\mathrm{DRAL}}}{\tau}
=\frac{1}{2}.
\label{eq:traj_dral}
\end{equation}
On HRVL and DRAL one has $\widetilde{W}=\mathrm{Re}(W)$ for all $\nu$, wherever
the weak value is defined. The exception is $\ket{A}$, which lies on the DRAL
path: there $\braket{D}{A}=0$ leaves $W$ undefined, while the measured
$\widetilde{W}$ remains equal to $1/2$. On HDVA the two coincide only in the
ideal limit $\nu=1$; for $\nu<1$ the measured $\widetilde{W}$ departs from
$\mathrm{Re}(W)$ and remains bounded near $\ket{A}$.
The HDVA trajectory has unequal weights in general and a nonzero
angle-dependent cross term, so it is sensitive to $\nu$. As $\nu$ increases from the projective value, the otherwise symmetric
$\cos^2\theta$ response becomes increasingly asymmetric about $\ket{V}$, with the
largest departure developing toward $\ket{A}$, where the ideal $\nu=1$ response
diverges; the minimum shifts just past $\ket{V}$ and dips slightly below zero
(Fig.~\ref{fig:theory_hdva}). In HRVL, the
coefficient of $\ket{V}$ is purely imaginary, so
$\mathrm{Re}(a\bar b)=0$ and the response is independent of
$\nu$. In DRAL, the weights remain equal; the cross term is present but
cancels between numerator and denominator in Eq. \eqref{eq:tmean}, giving
$\tau/2$ for all angles and all values of $\nu$.

\section{Experimental setup}
\label{sec:setup}

\begin{figure*}[htbp]
\centering
\includegraphics[width=0.92\textwidth]{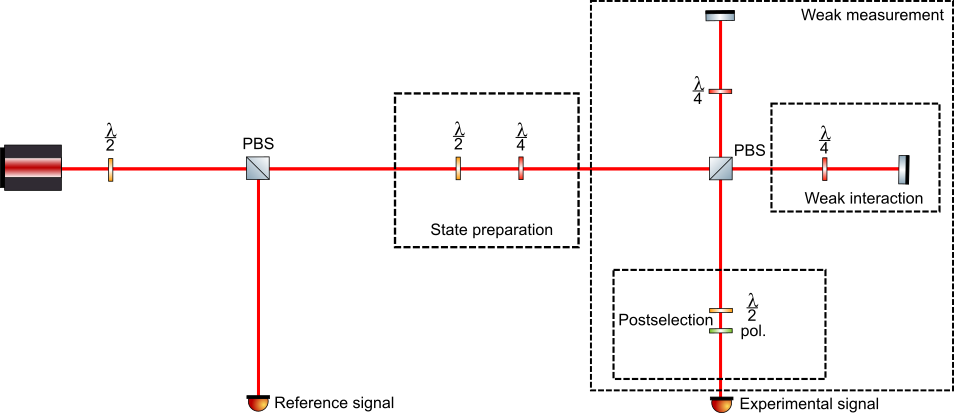}
\caption{Optical setup used to measure both quadratures of the weak value.
A half-wave plate (HWP, $\lambda/2$) and quarter-wave plate (QWP, $\lambda/4$)
in the state-preparation stage set the Poincar\'e-sphere trajectory. A
polarizing beam splitter (PBS) then separates the $\ket{H}$ and $\ket{V}$
components; the $\ket{H}$ arm is delayed before the paths recombine, and
post-selection on $\ket{D}$ is performed with an HWP and polarizer before
photodiode detection.}
\label{fig:real_setup}
\end{figure*}

A polarization qubit generated by a pulsed laser is weakly coupled to a temporal
pointer and post-selected in a polarization interferometer before detection.

\subsection{Pulsed laser source}
The experimental apparatus uses a commercial pulsed diode laser (Thorlabs
NPL64B \cite{ThorlabsNPL64B}) at $640~\mathrm{nm}$, chosen for accessibility and ease of integration
in a table-top optical setup. The pulse duration was set to
$10~\mathrm{ns}$ FWHM with a $10~\mathrm{MHz}$ repetition rate, producing a
near-flat-top temporal profile with a sharp leading edge. The sharp leading
edge is important because the time-shift extraction method in Sec.
\ref{sec:method} uses the maximum of the smoothed derivative rather than the
pulse peak. The laser is multimode, so the temporal pointer is only partially coherent at
the imposed delay. This visibility is obtained directly from the trajectory
data in Sec.~\ref{sec:results}, where it enters the partial-coherence model of
Eq. \eqref{eq:tmean}.

\subsection{Polarization interferometer}
The setup is shown in Fig. \ref{fig:real_setup}. The input qubit is
prepared from the $\ket{H}$ output of a polarizing beam splitter, followed by an
HWP and, for the HRVL and DRAL trajectories, a QWP set to
$\theta_{\mathrm{pol}}=0^\circ$ and $45^\circ$ respectively, producing the states
of Eqs.~\eqref{eq:state_hrvl} and \eqref{eq:state_dral}. The state-preparation
HWP is mounted on a motorized rotation stage (Thorlabs PRM1Z8 \cite{ThorlabsROT})
and is rotated to scan the polarization-state angle; the QWP, when present, is
fixed. A second polarizing beam splitter separates the $\ket{H}$ and $\ket{V}$
components into two arms; the $\ket{H}$ arm includes a round-trip optical
path-length difference of $\sim5~\mathrm{cm}$ before recombination on the same
beam splitter, implementing Eq. \eqref{eq:vonneumann} with a delay of order
$170~\mathrm{ps}$. The exact delay changes slightly between acquisitions because
of realignment and trigger drift, so the analysis uses the internal calibration
described in Sec. \ref{sec:method} rather than an externally imposed value of
$\tau$. The delay is small compared with the $10~\mathrm{ns}$ pulse duration but
large enough relative to the multimode coherence of the source to make $\nu$
experimentally relevant.

Post-selection on $\ket{D}$ is implemented with an HWP set to
$\theta_{\mathrm{pol}}=22.5^\circ$ followed by a linear polarizer set to
$\theta_{\mathrm{pol}}=0^\circ$. The post-selected signal is detected by a silicon photodiode
(Thorlabs DET025A, $2~\mathrm{GHz}$ bandwidth~\cite{ThorlabsDET025A}) and
recorded on an oscilloscope (Tektronix TDS5054B, $500~\mathrm{MHz}$ analog
bandwidth, $5~\mathrm{GS/s}$ real-time sampling rate~\cite{TektronixTDS5000}).
For the repetitive pulse train the oscilloscope is used in equivalent-time
sampling mode, which builds up the waveform to an effective spacing of one point
every $4~\mathrm{ps}$. The $500~\mathrm{MHz}$ analog bandwidth limits the recorded
rise time to several hundred picoseconds; the few-picosecond timing quoted below
is therefore not a resolution but the reproducibility with which the
leading-edge timing point (Sec.~\ref{sec:method}) is located, the bandwidth-limited
edge shape being common to all input states and cancelling in the internal
$\ket{H}/\ket{V}$ calibration.

\section{Temporal extraction method}
\label{sec:method}

\begin{figure}[htbp]
  \centering
  \includegraphics[width=\linewidth]{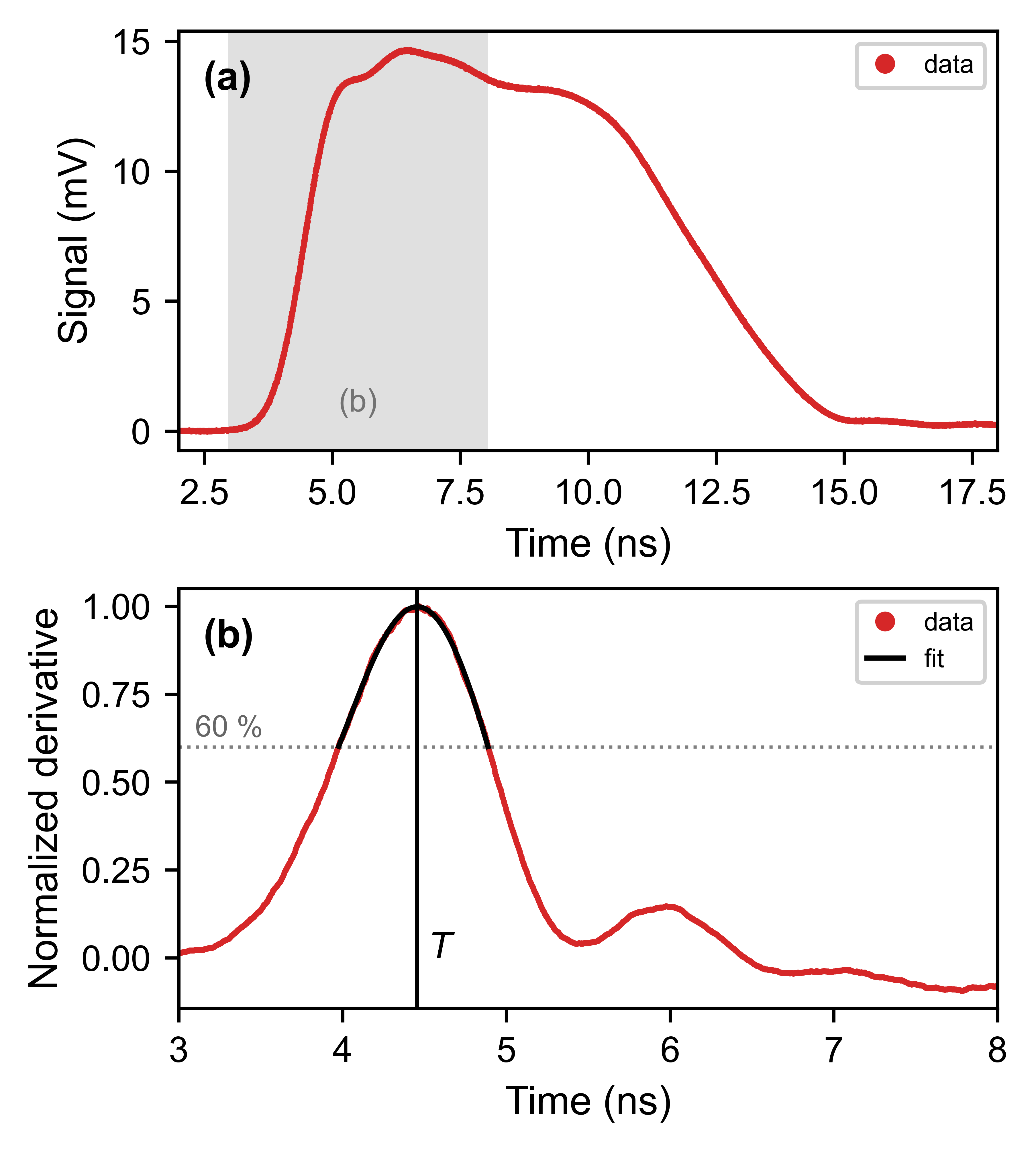}
  \caption{Time-shift extraction on a representative pulse.
    (a)~Recorded pulse $|F(t)|^2$, with its near-flat top and sharp leading
    edge; the shaded band marks the leading-edge region analyzed in~(b).
    (b)~Normalized smoothed first derivative $\mathrm{d}V/\mathrm{d}t$ of the
    leading edge (Savitzky--Golay filter). Points within $60\%$ of the maximum
    on both sides (dotted line) are retained for a fourth-order polynomial fit
    (solid black). The maximum of the fit defines the time $T$
    (vertical line).}
  \label{fig:extraction}
\end{figure}

We now describe how each pulse is reduced to a single time per trace and
combined with an $\ket{H}/\ket{V}$ calibration to obtain $\widetilde{W}$.

Each acquired pulse $|F(t)|^2$ has a near-flat top and a sharp leading edge (Fig.~\ref{fig:extraction}a), so
the most stable choice is the inflection point of the leading edge,
located by the maximum of the smoothed first derivative. This avoids the
ambiguity of peak-position or half-maximum methods on near-rectangular pulses.
First, the mean voltage over a pre-pulse window $t\in[0,3]~\mathrm{ns}$ is
subtracted from each trace to remove DC offsets and slow drifts. The first
derivative is then computed with a Savitzky--Golay filter using a 101-point
window, corresponding to $\sim404~\mathrm{ps}$, and a third-order
polynomial. The maximum of $dV/dt$ is located, and points within 60\% of that
maximum on both sides are retained for a local fit. A fourth-order polynomial
is fitted to the retained points, which is high enough to capture the slight
asymmetry of the derivative peak while remaining stable against the choice of
degree; using a second-order fit changes the extracted time $T$ by less than
$1~\mathrm{ps}$. The polynomial is evaluated on a $1000\times$ refined grid,
and the position of its maximum defines the time $T$ for that trace (Fig.~\ref{fig:extraction}b).

The HRVL trajectory passes through the $\ket{H}$ and $\ket{V}$ poles, and the
traces acquired at those poles provide the internal references used for
calibration. The references are defined by the pole they correspond to rather
than by a fixed HWP angle, since the absolute angle is not preserved across
acquisitions. The HDVA and DRAL trajectories do not sample these poles exactly
and are instead referenced to a separate $\ket{H}/\ket{V}$ acquisition. From these
references we extract $T_H$ and $T_V$ and define the session-specific delay
\begin{equation}
\tau_{\mathrm{int}}=T_H-T_V.
\label{eq:tau_int}
\end{equation}
The internal delay is typically $\sim177~\mathrm{ps}$. This internally measured quantity is what enters the data
analysis. Absolute trigger positions can drift by several hundred picoseconds
between acquisitions, whereas $T_H-T_V$ remains stable inside a single
measurement file. The post-selected normalized pointer shift $\widetilde{W}$ is therefore obtained as 
\begin{equation}
\widetilde{W}=\frac{T-T_V}{\tau_{\mathrm{int}}}.
\label{eq:wtilde}
\end{equation}
For the pure $\ket{H}$ and $\ket{V}$ references this calibration is exact by
translation of the pulse shape. For intermediate states $\widetilde{W}$ is given by Eq.~\eqref{eq:tmean} and
reduces to $\mathrm{Re}(W)$ only in the ideal limit $\nu=1$. The finite-coupling correction
is of order $(\tau/\sigma)^{2}\simeq2\times10^{-3}$ here, and is therefore
negligible compared with the visibility-induced departure; the deviations
reported in Sec.~\ref{sec:results} are attributed to $\nu$ alone. Across repeated acquisitions, the
dispersion of $T$ is typically a few picoseconds over five repetitions,
corresponding to a precision in $\widetilde{W}$ of order $0.02$. For each angle, the reported value is obtained from the average of the recorded
traces, and the associated uncertainty is the standard error of the per-trace
observables over the five repetitions.

The data-analysis method defined by the authors was implemented in Python,
using Anthropic Claude to assist with data consolidation and with translating
an earlier Matlab implementation into a batch-processing form. The numerical
method itself is that described above; the authors verified the implementation
and the data-to-angle assignments, and take responsibility for all reported
results.

\section{Results}
\label{sec:results}

We now report the measured real and imaginary quadratures along the three
trajectories, beginning with the scan and calibration procedure common to all
of them.

For each trajectory, the state-preparation HWP angle was scanned in
$2.5^\circ$ steps with five repetitions per angle, and
$\widetilde{W}$ was computed from Eq. \eqref{eq:wtilde}. The horizontal axis is the polarization-state angle $\theta$, obtained by
converting the HWP setting into the corresponding theoretical state angle, after
accounting for the offset between the HWP optical axis and the motor reading.
Each scan covers approximately $90^\circ$ of HWP rotation, corresponding to
$180^\circ$ in $\theta$. The HRVL and DRAL trajectories each use an additional QWP, set to
$\theta_{\mathrm{pol}}=0^\circ$ and $45^\circ$ respectively, while the HDVA
trajectory uses the HWP only. Because the HDVA and DRAL trajectories do not sample the $\ket{H}$ and $\ket{V}$
poles exactly, they cannot be calibrated internally; $T_V$ and the delay are
taken from a separate $\ket{H}/\ket{V}$ reference acquisition, which fixes only
the overall offset and scale and leaves the $\theta$-dependence untouched.

\subsection{HRVL trajectory: rotation about $S_2$}
\begin{figure}[htbp]
\centering
\includegraphics[width=\columnwidth]{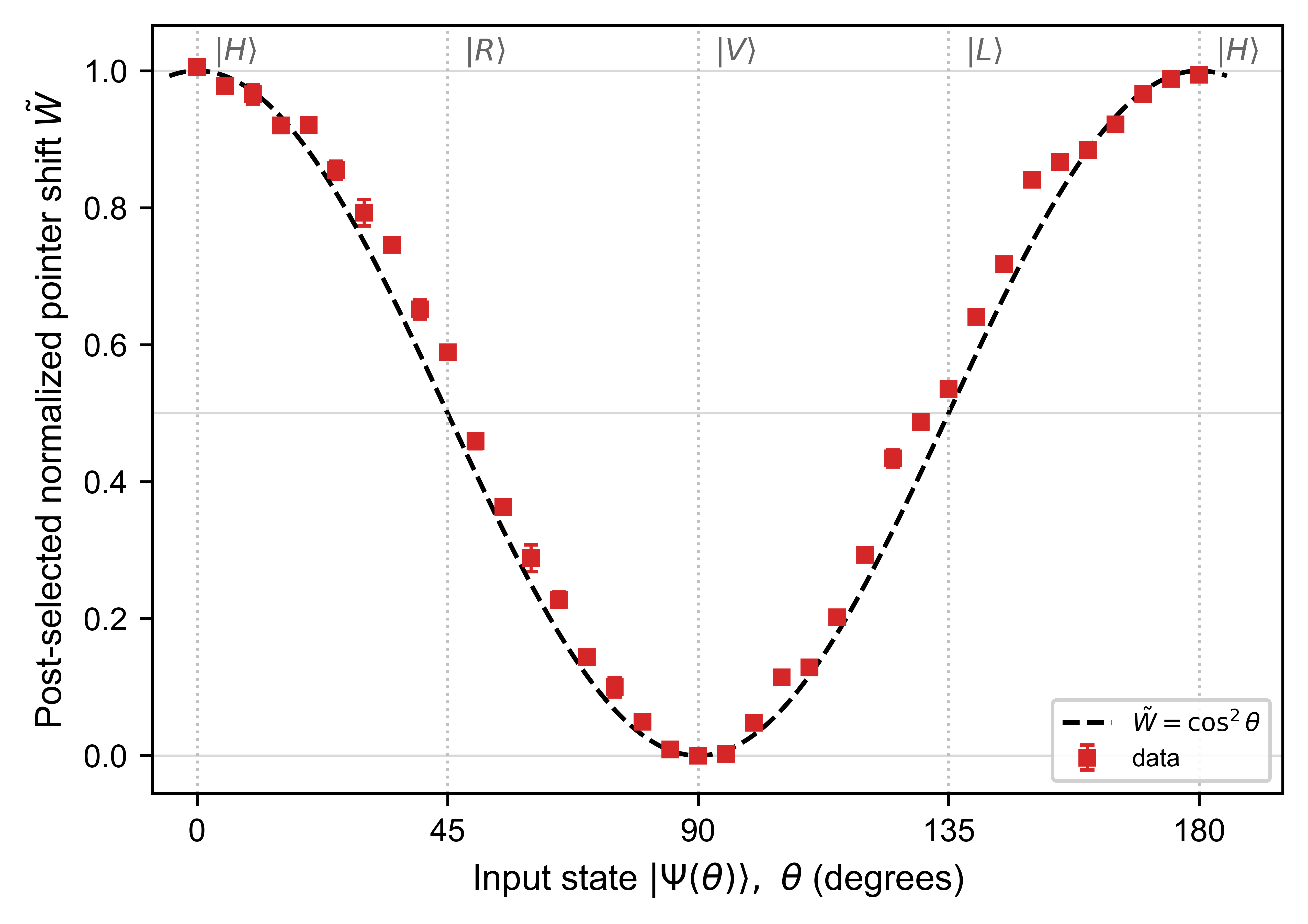}
\caption{Measured post-selected normalized pointer shift $\widetilde{W}$ along
the HRVL trajectory as a
function of the polarization-state angle $\theta$. Red squares: data obtained
from the internal $\ket{H}/\ket{V}$ calibration,
$\widetilde{W}=(T-T_V)/(T_H-T_V)$; error bars are the standard error over five
repetitions. Dashed line: $\cos^2\theta$, the parameter-free prediction for
this trajectory. The states $\ket{H}$, $\ket{R}$, $\ket{V}$, and $\ket{L}$ are
marked.}
\label{fig:hrvl_real}
\end{figure}

On the HRVL trajectory, the measured real part follows the parameter-free
prediction in Eq. \eqref{eq:traj_hrvl}, as shown in Fig. \ref{fig:hrvl_real}.
The expected response is
$\cos^2\theta$, and $\nu$ does not enter because
$ \mathrm{Re}(a\bar b)=0$ by construction. The data follow this prediction over
the full trajectory from $\ket{H}$ through $\ket{R}$, $\ket{V}$, and
$\ket{L}$ back toward $\ket{H}$. The RMS residual is $\sim0.04$, which
is consistent with the timing precision obtained from the internal $\ket{H}/\ket{V}$
calibration. This agreement without any fitted parameter confirms that the
time-shift extraction method reproduces the visibility-independent response of
this trajectory.

\subsection{DRAL trajectory: rotation about $S_1$}
\begin{figure}[htbp]
\centering
\includegraphics[width=\columnwidth]{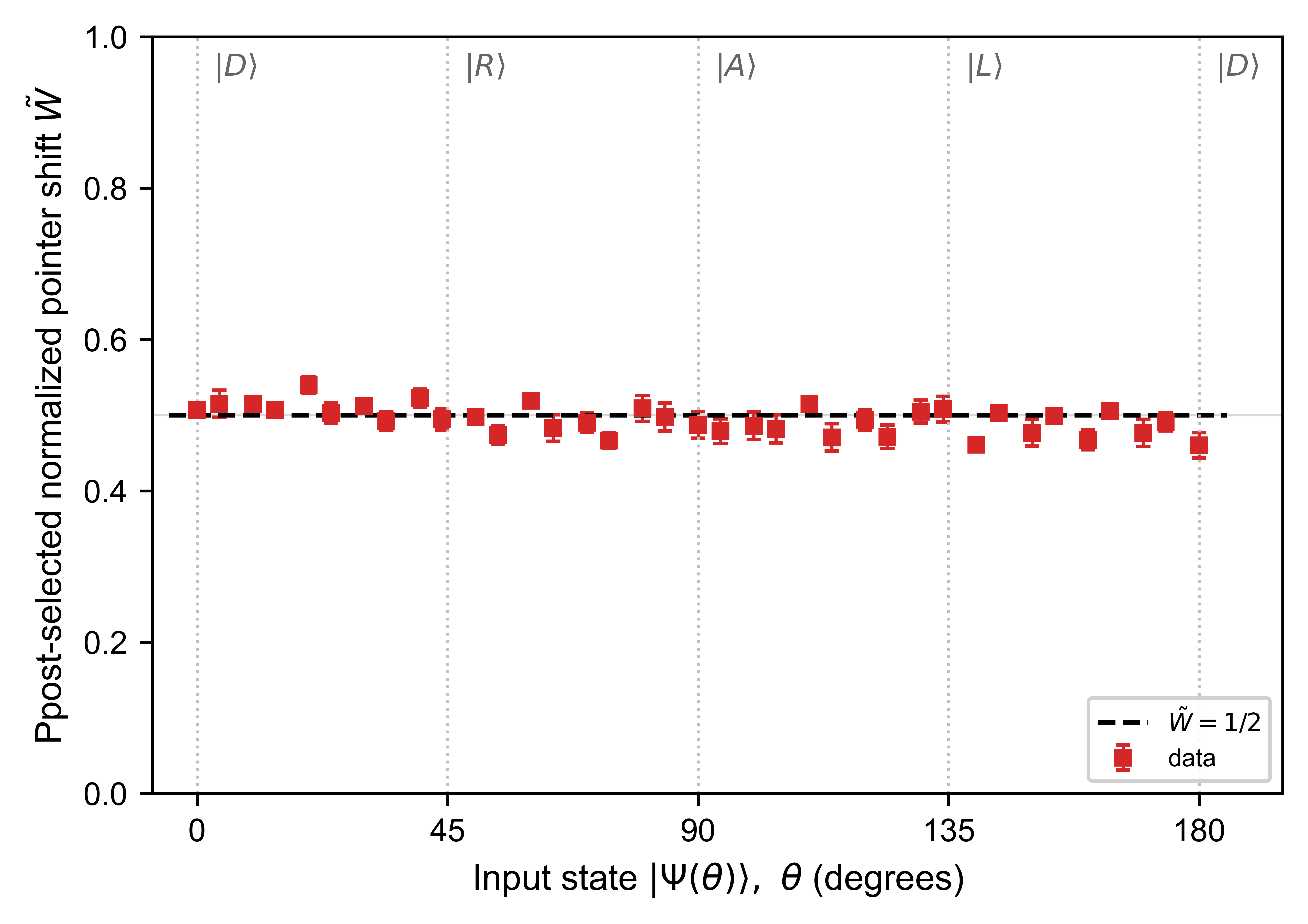}
\caption{Measured post-selected normalized pointer shift $\widetilde{W}$ along
the DRAL trajectory as a
function of the polarization-state angle $\theta$. Red squares: data; error
bars are the standard error over five repetitions. Dashed line: the
parameter-free prediction $\widetilde{W}=1/2$. Because the DRAL trajectory does
not pass through the $\ket{H}$ and $\ket{V}$ poles, $T_V$ and the delay are
taken from a separate $\ket{H}/\ket{V}$ reference acquisition rather than from
the scan itself; the calibration-independent result is the constancy of the
response. The mean value is $0.50\pm0.01$.}
\label{fig:dral_real}
\end{figure}

The DRAL trajectory tests the second visibility-independent prediction,
Eq. \eqref{eq:traj_dral}, as shown in Fig. \ref{fig:dral_real}. Along this
path the expected time shift is constant, with $\widetilde{W}=1/2$ for all
angles. Because the DRAL scan does not cross the $\ket{H}$ and $\ket{V}$ poles,
$T_V$ and the delay are taken from a separate $\ket{H}/\ket{V}$ reference
acquisition; this fixes only the overall offset and scale and leaves the
$\theta$-dependence untouched, so the test of constancy is independent of the
calibration chosen. The measured response is constant across the trajectory,
with a mean level of $0.50\pm0.01$. This confirms the cancellation of the
$\nu$-dependent terms between the numerator and denominator of
Eq. \eqref{eq:tmean}.

\subsection{HDVA trajectory: rotation about $S_3$}
\begin{figure}[htbp]
\centering
\includegraphics[width=\columnwidth]{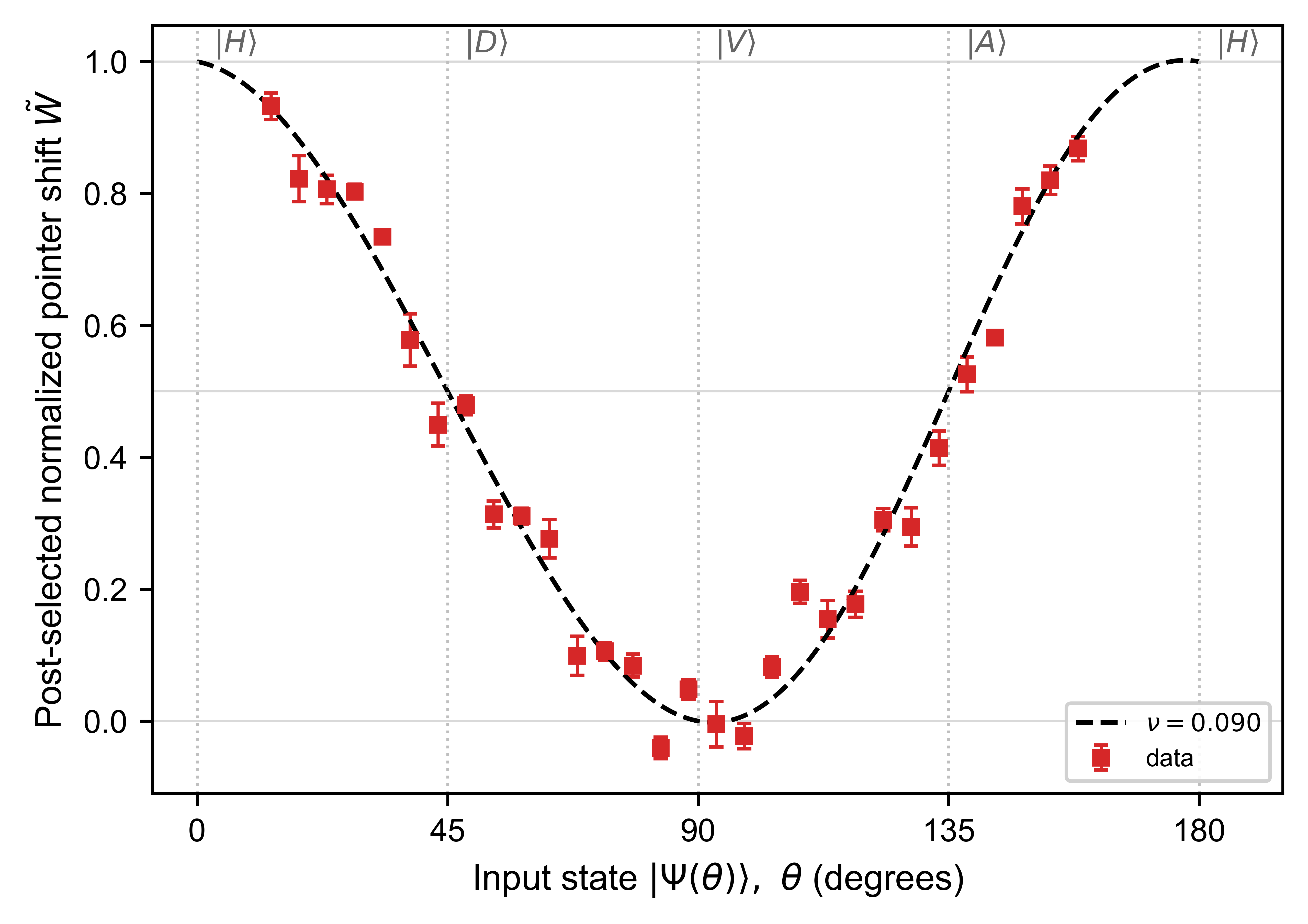}
\caption{Measured post-selected normalized pointer shift $\widetilde{W}$ along
the HDVA trajectory as a
function of the polarization-state angle $\theta$. Red squares: data; error
bars are the standard error over five repetitions. Solid line: single-parameter
fit to the partial-coherence model of Eq.~\eqref{eq:traj_hdva}. Since the scan
does not sample $\ket{V}$ exactly, $T_V$ and the delay are taken from a separate
reference acquisition. The model reproduces the measured response. The fitted
visibility is $\nu=0.090\pm0.05$.}
\label{fig:hdva_real}
\end{figure}

The HDVA trajectory is sensitive to partial pointer coherence and therefore
tests the full model in Eq. \eqref{eq:traj_hdva}, as shown in
Fig. \ref{fig:hdva_real}. The measured response is non-monotonic and,
consistent with Fig.~\ref{fig:theory_hdva}, asymmetric about $\ket{V}$ with the
largest excursion toward $\ket{A}$, showing that the partial coherence of the
pointer has an observable effect. Fitting Eq. \eqref{eq:traj_hdva} with $\nu$ as
the only free parameter gives $\nu=0.09\pm0.05$, indicating a pointer of reduced
coherence well below the ideal limit $\nu=1$. The uncertainty is dominated by
the residual scatter of this trajectory, in which the visibility enters only
through the $\sin4\theta$ term of amplitude $\nu/4$; this sets the sensitivity
of the method at $\delta\nu=4\sigma\sqrt{2/N}$, where $N$ is the number of
angular points on the trajectory and $\sigma$ the residual scatter in
$\widetilde{W}$. The reduced visibility is
consistent with the use of a multimode pulsed source whose coherence length is
short compared with the imposed delay. Near $\ket{A}$, where the ideal coherent
weak value would diverge, the measured response remains finite, supporting the
partial-coherence description. Together with the parameter-free HRVL and DRAL
tests, the HDVA fit shows that a single visibility parameter describes the
real-quadrature behavior of the partially coherent pointer.

\subsection{Imaginary quadrature}
\begin{figure}[htbp]
\centering
\includegraphics[width=\columnwidth]{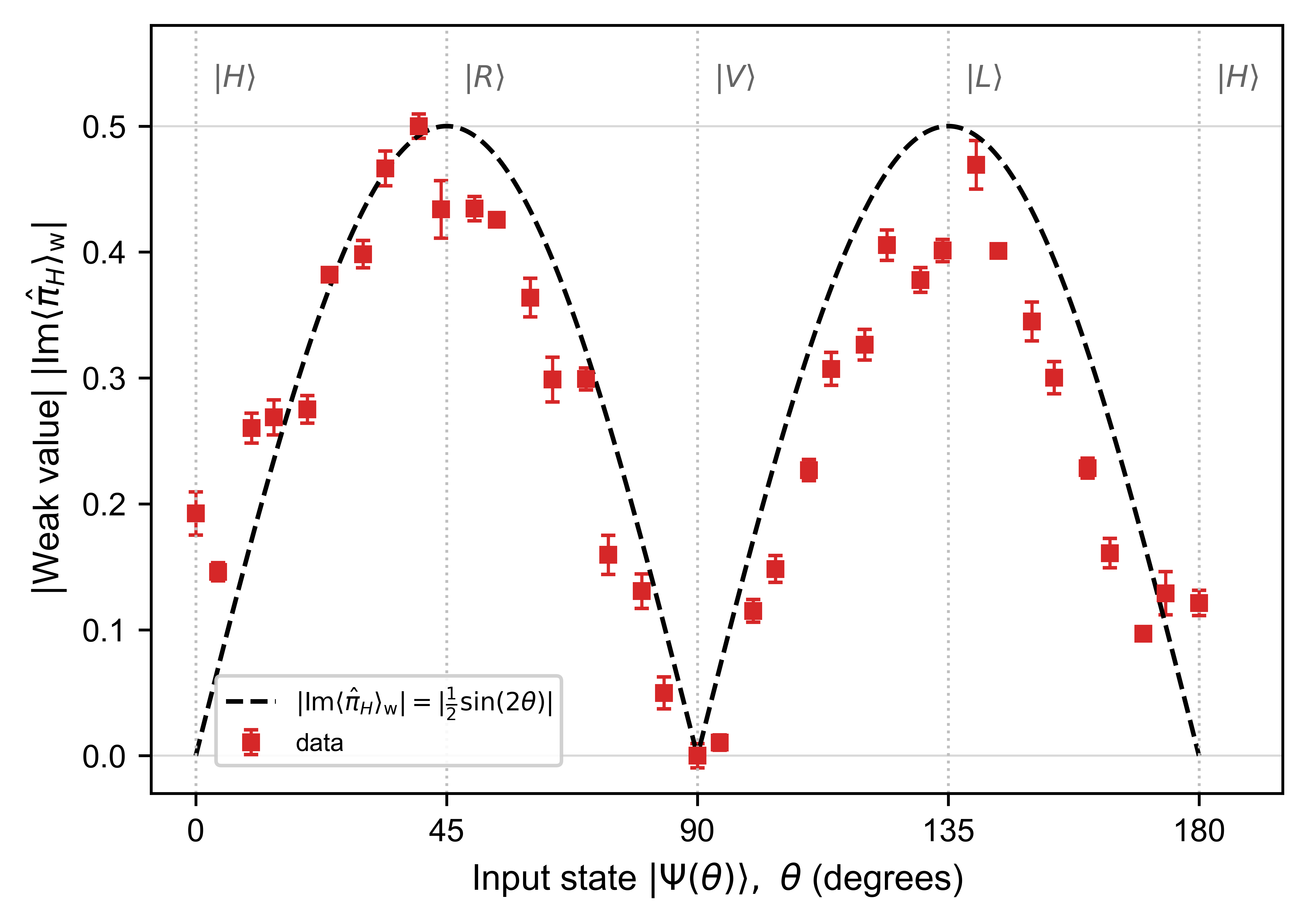}
\caption{Imaginary part of the weak value, $|\mathrm{Im}(W)|$, along the HRVL
trajectory as a function of the polarization-state angle $\theta$. Red squares:
amplitude obtained from the shift of the frequency centroid of the full Fourier
transform of the recorded signal; error bars are the standard error over five
repetitions. Dashed line: the predicted $|\tfrac12\sin(2\theta)|$ shape,
normalized to its maximum. Because the centroid is computed from a positive
power spectrum, the method accesses the magnitude of $\mathrm{Im}(W)$ but not
its sign.}
\label{fig:imag_results}
\end{figure}

The same temporal traces also give qualitative access to the imaginary part of
the weak value through the spectral content of the recorded pointer. For each
input state, the power spectral density of the recorded signal is computed from
the full Fourier transform, and its centroid frequency is evaluated as
\begin{equation}
f_c=\frac{\sum_i f_i P_i}{\sum_i P_i}.
\label{eq:spectral_centroid}
\end{equation}
The $\ket{V}$ state is used as the zero reference, as it corresponds to zero
centroid shift, consistent with $\mathrm{Im}(W)$ vanishing there. The centroid shift follows the expected
$|\sin(2\theta)|$ dependence along the HRVL trajectory, with maxima at the
circular states $\ket{R}$ and $\ket{L}$ and a minimum at $\ket{V}$, as shown in
Fig. \ref{fig:imag_results}. This is presented as a qualitative demonstration,
not a calibrated measurement of $\mathrm{Im}(W)$. Equation~\eqref{eq:omega_shift}
predicts a frequency shift proportional to $\mathrm{Im}(W)$ for an ideal coherent
pointer. With the partially coherent multimode pointer used here, the readout
instead accesses only the magnitude: the spectral measurement of
the imaginary part can be understood as a self-homodyne detection in the
incoherent regime, where the two polarization components act as signal and
delayed reference~\cite{hui2009}. Because the photocurrent spectrum is centered
at zero intermediate frequency and folded about it, the method recovers the
magnitude of the imaginary weak value but not its sign. The sign could be
retrieved with a self-heterodyne scheme, in which a frequency shifter moves the
intermediate frequency away from zero~\cite{hui2009}. The residual scatter between neighboring states is larger than the per-state
statistical uncertainty, consistent with the slow drifts between acquisitions
discussed above.

\section{Discussion and Conclusion}
\label{sec:discussion}
The three Poincar\'e-axis trajectories exhibit three distinct responses to
partial pointer coherence, all consistent with Eq.~\eqref{eq:tmean}. HRVL and
DRAL are parameter-free checks: HRVL follows $\cos^2\theta$ with an RMS
residual of $\sim0.04$, while DRAL remains constant at $0.50\pm0.01$. The
residual scatter follows the same hierarchy. DRAL is constant in $\theta$ and
independent of $\nu$, so it is insensitive both to the angle calibration and to
drifts of the source coherence; it is correspondingly the tightest of the three.
HRVL varies with $\theta$ but not with $\nu$, while the HDVA trajectory below
depends on both, and their residuals order accordingly. The uncertainty quoted
for each point is the statistical error over five repetitions and does not
include the angle assignment, which is common to a whole scan. The HDVA
trajectory is captured by the single-parameter partial-coherence model with
$\nu=0.09\pm0.05$. This reduced visibility reflects the short coherence length
of the multimode source relative to the $\sim177~\mathrm{ps}$ delay, consistent
with a pointer well below the ideal weak limit. The calibration procedure provides
few-picosecond timing precision on $10~\mathrm{ns}$ pulses.

The main limitation of the present experiment is the imaginary-quadrature
readout. The frequency-centroid method recovers the angular dependence and
magnitude of the imaginary part, but it loses the sign because it is computed
from a positive power spectrum. Phase-preserving detection, for example
heterodyne or interferometric spectral readout, would remove this ambiguity.
The second limitation is that the visibility parameter compresses a multimode
pointer structure into a single number; this is sufficient at a fixed delay but
would need refinement for scans across the full multimode beating structure.
The experiment nevertheless shows that partial-coherence weak measurement is
accessible with commercial pulsed sources and fast oscilloscopes.

We have measured the post-selected normalized pointer shift across three
Poincar\'e-sphere trajectories in the partial-coherence regime, reaching
few-picosecond precision on nanosecond pulses. The three trajectories are rotations about the principal
axes of the Poincar\'e sphere and, taken together, realize the distinct
responses of the time shift to pointer coherence: HRVL carries no interference
cross term and DRAL carries one that cancels between numerator and denominator,
so both are parameter-free and independent of $\nu$ by construction, while HDVA
retains the cross term and constrains the visibility. The two $\nu$-independent
trajectories are therefore not an incidental check but a control imposed by the
geometry; their agreement with the parameter-free predictions, together with the
single-parameter HDVA fit, tests the partial-coherence model without any
external calibration of $\nu$.

A more coherent pulsed source would extend the measurement toward the ideal weak
limit $\nu=1$, where $\widetilde{W}$ recovers $\mathrm{Re}(W)$. Phase-preserving
spectral detection, such as a self-heterodyne readout, would restore the sign of
the imaginary quadrature; together with a more coherent source and the
post-selected geometry used here, which realizes the projector weak value
$W=a/(a+b)$, this would turn the trajectory scans into a direct route to
reconstructing the input polarization
state~\cite{Lundeen_Direct_Measurement,Guilleaum}. More broadly, a temporal
pointer requires only a delay and time- or frequency-resolved detection, both
native to integrated photonics, so the scheme is in principle more amenable to
on-chip implementation than spatial-pointer measurements that rely on free-space
propagation.

\section*{CRediT author statement}
S. P. Gervais: Conceptualization, Investigation, Methodology, Software,
Formal analysis, Data curation, Writing -- original draft.
L. Giner: Conceptualization, Methodology, Formal analysis, Data curation,
Resources, Supervision, Project administration, Validation, Visualization,
Writing -- original draft, Writing -- review and editing, Funding acquisition.

\section*{Declaration of generative AI and AI-assisted technologies in the
manuscript preparation process}
During the preparation of this work the authors used Anthropic Claude for
language editing, manuscript-structure review, and LaTeX formatting
assistance. After using this tool, the authors reviewed and edited the
content as needed and take full responsibility for the content of the
published article.

\section*{Declaration of competing interest}
The authors declare that they have no known competing financial interests or
personal relationships that could have appeared to influence the work reported
in this paper.

\section*{Data availability}
Data supporting the findings of this study are available from the corresponding
author upon reasonable request.

\section*{Funding}
This work was supported by the Natural Sciences and Engineering Research
Council of Canada (RGPIN-2022-04394); the Canada Foundation for Innovation
(John R. Evans Leaders Fund 44031); the New Brunswick Innovation Foundation
(Start-up Awards, Equipment Fund EQP\_2024\_016); the Universit\'e de Moncton
(Facult\'e des Sciences and Facult\'e des \'Etudes Sup\'erieures); and Mitacs
through the Globalink Research Internship program.

\section*{Acknowledgements}
The authors thank S. Nandja and C. Ndiaye for contributions to the theoretical
modeling of weak measurements with a temporal pointer, M. Tardat and T. Blandin
for theoretical and preliminary experimental work on pulsed interferometry, and
E. Wenzel and H. Heineman for assistance with instrument interfacing and the
fast data-acquisition system.

\bibliographystyle{elsarticle-num}
\bibliography{references}

@article{Aharonov,
  title = {How the result of a measurement of a component of the spin of a spin-1/2 particle can turn out to be 100},
  author = {Aharonov, Yakir and Albert, David Z. and Vaidman, Lev},
  journal = {Phys. Rev. Lett.},
  volume = {60},
  issue = {14},
  pages = {1351--1354},
  numpages = {0},
  year = {1988},
  month = {Apr},
  publisher = {American Physical Society},
  doi = {10.1103/PhysRevLett.60.1351}
}

@book{JordanBook,
  author         = {Andrew N. Jordan and Irfan A. Siddiqi},
  editor         = {1},
  publisher      = {Cambridge University Press},
  title          = {Quantum Measurement: Theory and Practice},
  year           = {2024}
}

@article{Lundeen_Bamber,
  title = {Procedure for Direct Measurement of General Quantum States Using Weak Measurement},
  author = {Lundeen, Jeff S. and Bamber, Charles},
  journal = {Phys. Rev. Lett.},
  volume = {108},
  issue = {7},
  pages = {070402},
  numpages = {5},
  year = {2012},
  month = {Feb},
  publisher = {American Physical Society},
  doi = {10.1103/PhysRevLett.108.070402}
}

@article{Lundeen_Resch,
title = {Practical measurement of joint weak values and their connection to the annihilation operator},
journal = {Physics Letters A},
volume = {334},
number = {5},
pages = {337-344},
year = {2005},
issn = {0375-9601},
doi = {https://doi.org/10.1016/j.physleta.2004.11.037},
author = {J.S. Lundeen and K.J. Resch}
}

@article{Guilleaum,
  title = {Direct Measurement of the Density Matrix of a Quantum System},
  author = {Thekkadath, G. S. and Giner, L. and Chalich, Y. and Horton, M. J. and Banker, J. and Lundeen, J. S.},
  journal = {Phys. Rev. Lett.},
  volume = {117},
  issue = {12},
  pages = {120401},
  numpages = {5},
  year = {2016},
  month = {Sep},
  publisher = {American Physical Society},
  doi = {10.1103/PhysRevLett.117.120401}
}

@article{WeakorStd,
  title = {Measuring Small Longitudinal Phase Shifts: Weak Measurements or Standard Interferometry?},
  author = {Brunner, Nicolas and Simon, Christoph},
  journal = {Phys. Rev. Lett.},
  volume = {105},
  issue = {1},
  pages = {010405},
  numpages = {4},
  year = {2010},
  month = {Jul},
  publisher = {American Physical Society},
  doi = {10.1103/PhysRevLett.105.010405}
}

@article{Hairiri,
  title = {Experimental simultaneous readout of the real and imaginary parts of the weak value},
  author = {Hariri, A. and Curic, D. and Giner, L. and Lundeen, J. S.},
  journal = {Phys. Rev. A},
  volume = {100},
  issue = {3},
  pages = {032119},
  numpages = {7},
  year = {2019},
  month = {Sep},
  publisher = {American Physical Society},
  doi = {10.1103/PhysRevA.100.032119}
}

@article{OpticalNetworks,
  title = {Optical Telecom Networks as Weak Quantum Measurements with Postselection},
  author = {Brunner, Nicolas and Ac\'{\i}n, Antonio and Collins, Daniel and Gisin, Nicolas and Scarani, Valerio},
  journal = {Phys. Rev. Lett.},
  volume = {91},
  issue = {18},
  pages = {180402},
  numpages = {4},
  year = {2003},
  month = {Oct},
  publisher = {American Physical Society},
  doi = {10.1103/PhysRevLett.91.180402}
}

@article{Brunner_2004,
   title={Direct Measurement of Superluminal Group Velocity and Signal Velocity in an Optical Fiber},
   volume={93},
   ISSN={1079-7114},
   doi={10.1103/physrevlett.93.203902},
   number={20},
   journal={Physical Review Letters},
   publisher={American Physical Society (APS)},
   author={Brunner, Nicolas and Scarani, Valerio and Wegmüller, Mark and Legré, Matthieu and Gisin, Nicolas},
   year={2004},
   month={Nov} 
}

@article{HardyParadox,
  title = {Quantum mechanics, local realistic theories, and Lorentz-invariant realistic theories},
  author = {Hardy, Lucien},
  journal = {Phys. Rev. Lett.},
  volume = {68},
  issue = {20},
  pages = {2981--2984},
  numpages = {0},
  year = {1992},
  month = {May},
  publisher = {American Physical Society},
  doi = {10.1103/PhysRevLett.68.2981}
}

@article{Aharonov_2002_Hardy,
   title={Revisiting Hardy’s paradox: counterfactual statements, real measurements, entanglement and weak values},
   volume={301},
   ISSN={0375-9601},
   DOI={10.1016/s0375-9601(02)00986-6},
   number={3–4},
   journal={Physics Letters A},
   publisher={Elsevier BV},
   author={Aharonov, Yakir and Botero, Alonso and Popescu, Sandu and Reznik, Benni and Tollaksen, Jeff},
   year={2002},
   month=aug, pages={130–138} 
}

@article{Btwn_pre_post,
title = {What is the value of an observable between pre- and postselection?},
journal = {Physics Letters A},
volume = {322},
number = {5},
pages = {298-300},
year = {2004},
issn = {0375-9601},
doi = {https://doi.org/10.1016/j.physleta.2004.01.041},
author = {Lars M Johansen}
}

@article{Lundeen_Direct_Measurement,
	author = {Lundeen, Jeff S. and Sutherland, Brandon and Patel, Aabid and Stewart, Corey and Bamber, Charles},
	date = {2011/06/01},
	doi = {10.1038/nature10120},
	id = {Lundeen2011},
	isbn = {1476-4687},
	journal = {Nature},
	number = {7350},
	pages = {188--191},
	title = {Direct measurement of the quantum wavefunction},
	volume = {474},
	year = {2011}
}

@article{weak_photon_pol,
  title = {Measurement of Quantum Weak Values of Photon Polarization},
  author = {Pryde, G. J. and O'Brien, J. L. and White, A. G. and Ralph, T. C. and Wiseman, H. M.},
  journal = {Phys. Rev. Lett.},
  volume = {94},
  issue = {22},
  pages = {220405},
  numpages = {4},
  year = {2005},
  month = {Jun},
  publisher = {American Physical Society},
  doi = {10.1103/PhysRevLett.94.220405}
}

@article{Magaña-Loaiza_2017,
doi = {10.1088/1402-4896/92/2/023001},
year = {2016},
month = {dec},
publisher = {IOP Publishing},
volume = {92},
number = {2},
pages = {023001},
author = {Omar S Magaña-Loaiza and Jérémie Harris and Jeff S Lundeen and Robert W Boyd},
title = {Weak-value measurements can outperform conventional measurements},
journal = {Physica Scripta}
}

@article{Salazar,
  title = {Measurement of sub-pulse-width temporal delays via spectral interference induced by weak value amplification},
  author = {Salazar-Serrano, Luis Jos\'e and Janner, Davide and Brunner, Nicolas and Pruneri, Valerio and Torres, Juan P.},
  journal = {Phys. Rev. A},
  volume = {89},
  issue = {1},
  pages = {012126},
  numpages = {5},
  year = {2014},
  month = {Jan},
  publisher = {American Physical Society},
  doi = {10.1103/PhysRevA.89.012126}
}

@article{Jeff_outperform,
  title = {Weak Value Amplification Can Outperform Conventional Measurement in the Presence of Detector Saturation},
  author = {Harris, J\'er\'emie and Boyd, Robert W. and Lundeen, Jeff S.},
  journal = {Phys. Rev. Lett.},
  volume = {118},
  issue = {7},
  pages = {070802},
  numpages = {6},
  year = {2017},
  month = {Feb},
  publisher = {American Physical Society},
  doi = {10.1103/PhysRevLett.118.070802}
}

@article{Howell2022,
  title = {Weakness of weak values: Incompatibility of anomalous pulse-spectrum amplification and optical frequency combs},
  author = {Howell, John C.},
  journal = {Physical Review A},
  volume = {106},
  number = {1},
  pages = {012224},
  year = {2022},
  month = {Jul},
  publisher = {American Physical Society},
  doi = {10.1103/PhysRevA.106.012224}
}

@manual{TektronixTDS5000,
  title        = {TDS5000 Series Digital Phosphor Oscilloscope User Manual},
  author       = {{Tektronix, Inc.}},
  year         = {2002},
  organization = {Tektronix, Inc.},
  address      = {Beaverton, OR, USA},
  note         = {Document Number: 071-0858-05},
  url          = {https://www.tek.com/en/oscilloscope/tds5052-manual/tds5000-series-digital-phosphor-oscilloscopes}
}

@article{ApplicationWeak,
  title = {Foundations and applications of weak quantum measurements},
  author = {Aharonov, Yakir and Cohen, Eliahu and Elitzur, Avshalom C.},
  journal = {Phys. Rev. A},
  volume = {89},
  issue = {5},
  pages = {052105},
  numpages = {10},
  year = {2014},
  month = {May},
  publisher = {American Physical Society},
  doi = {10.1103/PhysRevA.89.052105}
}

@misc{ThorlabsNPL64B,
  author       = {Thorlabs},
  title        = {NPL64B - Nanosecond Pulsed Laser Diode System, 640 nm, 5 - 39 ns Adjustable Pulse Width},
  howpublished = {\url{https://www.thorlabs.com/item/NPL64B}},
  note         = {Accessed: 20 June 2026}
}

@misc{ThorlabsDET025A,
  author       = {Thorlabs},
  title        = {DET025A - 2 GHz Si Free-Space Photodetector with Window, 400 - 1100 nm, 8-32 Tap},
  howpublished = {\url{https://www.thorlabs.com/item/DET025A}},
  note         = {Accessed: 20 June 2026}
}

@misc{ThorlabsROT,
  author       = {Thorlabs},
  title        = {PRM1Z8 - {\O}1" Motorized Precision Rotation Stage},
  howpublished = {\url{https://www.thorlabs.com/item/PRM1Z8}},
  note         = {Accessed: 20 June 2026}
}

@article{Time_delay_china,
  title = {Precision estimation of time delay based on weak measurement with real weak value},
  author = {Liu, Weining and Wang, Yisen and Luo, Hailu},
  journal = {Phys. Rev. A},
  volume = {108},
  issue = {5},
  pages = {052606},
  numpages = {7},
  year = {2023},
  month = {Nov},
  publisher = {American Physical Society},
  doi = {10.1103/PhysRevA.108.052606}
}

@misc{troupe2017quantumcryptographyweakmeasurements,
      title={Quantum Cryptography with Weak Measurements}, 
      author={James E. Troupe and Jacob M. Farinholt},
      year={2017},
      eprint={1702.04836},
      archivePrefix={arXiv},
      primaryClass={quant-ph},
      url={https://arxiv.org/abs/1702.04836}, 
}

@article{imaginary_part,
  title = {Direct experimental test of commutation relation via imaginary weak value},
  author = {Wagner, Richard and Kersten, Wenzel and Danner, Armin and Lemmel, Hartmut and Pan, Alok Kumar and Sponar, Stephan},
  journal = {Phys. Rev. Res.},
  volume = {3},
  issue = {2},
  pages = {023243},
  numpages = {7},
  year = {2021},
  month = {Jun},
  publisher = {American Physical Society},
  doi = {10.1103/PhysRevResearch.3.023243}
}

@article{quantuminfoweak,
  title = {Interferometric weak value deflections: Quantum and classical treatments},
  author = {Howell, John C. and Starling, David J. and Dixon, P. Ben and Vudyasetu, Praveen K. and Jordan, Andrew N.},
  journal = {Phys. Rev. A},
  volume = {81},
  issue = {3},
  pages = {033813},
  numpages = {3},
  year = {2010},
  month = {Mar},
  publisher = {American Physical Society},
  doi = {10.1103/PhysRevA.81.033813}
}

@article{intlundeen,
  title = {Enhancing Interferometry Using Weak Value Amplification with Real Weak Values},
  author = {Huang, Jing-Hui and Jordan, Kyle M. and Dada, Adetunmise C. and Hu, Xiang-Yun and Lundeen, Jeff. S.},
  journal = {Phys. Rev. Lett.},
  volume = {134},
  issue = {8},
  pages = {080802},
  numpages = {6},
  year = {2025},
  month = {Feb},
  publisher = {American Physical Society},
  doi = {10.1103/PhysRevLett.134.080802}
}

@article{QuantumLimitedMetrology,
  title = {Approaching Quantum-Limited Metrology with Imperfect Detectors by Using Weak-Value Amplification},
  author = {Xu, Liang and Liu, Zexuan and Datta, Animesh and Knee, George C. and Lundeen, Jeff S. and Lu, Yan-qing and Zhang, Lijian},
  journal = {Phys. Rev. Lett.},
  volume = {125},
  issue = {8},
  pages = {080501},
  numpages = {6},
  year = {2020},
  month = {Aug},
  publisher = {American Physical Society},
  doi = {10.1103/PhysRevLett.125.080501}
}

@article{RydbergDispersion,
	author = {Jiang, Yinghang and Wu, Jiguo and Shi, Meng and Zheng, Hanqing and Guo, Fei and Xiao, Zhiguang and Zhang, Zhiyou},
	date = {2025/04/09},
	doi = {10.1038/s42005-025-02014-3},
	id = {Jiang2025},
	isbn = {2399-3650},
	journal = {Communications Physics},
	number = {1},
	pages = {144},
	title = {Quantum weak measurement amplifies dispersion signal of Rydberg atomic system},
	volume = {8},
	year = {2025}
  }

@article{Huang:22,
author = {Jing-Hui Huang and Xue-Ying Duan and Guang-Jun Wang and Xiang-Yun Hu},
journal = {J. Opt. Soc. Am. B},
number = {4},
pages = {1289--1296},
publisher = {Optica Publishing Group},
title = {Enhancing the precision of detecting weak magnetic fields based on weak-value amplification},
volume = {39},
month = {Apr},
year = {2022},
doi = {10.1364/JOSAB.447515}
}

@article{SpinHallEffect,
author = {Onur Hosten  and Paul Kwiat },
title = {Observation of the Spin Hall Effect of Light via Weak Measurements},
journal = {Science},
volume = {319},
number = {5864},
pages = {787-790},
year = {2008},
doi = {10.1126/science.1152697},
eprint = {https://www.science.org/doi/pdf/10.1126/science.1152697}}

@incollection{hui2009,
  author    = {Hui, Rongqing and O'Sullivan, Maurice},
  title     = {Characterization of optical devices},
  booktitle = {Fiber Optic Measurement Techniques},
  publisher = {Academic Press},
  year      = {2009},
  chapter   = {3},
  pages     = {259--363},
  isbn      = {9780123738653},
  doi       = {10.1016/B978-0-12-373865-3.00003-3}
}

@article{Cho2010,
  title = {Weak value measurement with an incoherent measuring device},
  author = {Cho, Young-Wook and Lim, Hyang-Tag and Ra, Young-Sik and Kim, Yoon-Ho},
  journal = {New Journal of Physics},
  volume = {12},
  number = {2},
  pages = {023036},
  year = {2010},
  doi = {10.1088/1367-2630/12/2/023036}
}

@article{Jozsa2007,
  title = {Complex weak values in quantum measurement},
  author = {Jozsa, Richard},
  journal = {Phys. Rev. A},
  volume = {76},
  issue = {4},
  pages = {044103},
  year = {2007},
  doi = {10.1103/PhysRevA.76.044103}
}

\end{document}